\documentclass[a4paper, 12pt]{article}

\usepackage[margin=1.2in]{geometry}

\newcommand{\address}[1]{\gdef\myaddress{#1}}
\newcommand{\keywords}[1]{\gdef\mykeywords{#1}}
\renewcommand{\abstract}[1]{\gdef\myabstract{#1}}

\title{Generalised exponential families and associated
entropy functions}

\author{Jan Naudts}
\date {March 2008}
\address{
Departement Fysica, Universiteit Antwerpen, \\
Groenenborgerlaan 171, 2020 Antwerpen, Belgium\\
E-mail: jan.naudts@ua.ac.be \\ }

\abstract{
A generalised notion of exponential families is introduced. It is
based on the variational principle, borrowed from statistical physics.
It is shown that inequivalent generalised entropy functions
lead to distinct generalised exponential families.
The well-known result that the inequality of Cram\'er and Rao
becomes an equality in the case of an exponential family can be
generalised. However, this requires the introduction of escort probabilities.
}


\newcommand{\be}{\begin{eqnarray}}
\newcommand{\ee}{\end{eqnarray}}
\newtheorem{definition}{Definition}
\newtheorem{proposition}{Proposition}
\newtheorem{lemma}{Lemma}
\newtheorem{theorem}{Theorem}
\def\beginproof{\par\strut\vskip 0.01cm\noindent{\bf Proof}\par\noindent}
\def\endproof{\par\strut\hfill$\square$\par\vskip 0.5cm}
\def\Tr {\,{\rm Tr}\,}

\usepackage{amsfonts}
\def\Ro{{\mathbb R}}

\begin{document}
\maketitle
\centerline {\parbox{10 cm}{\myaddress}}

\paragraph{Abstract}
\myabstract

%%%%%%%%%%%%%%%%%%%%%%%%%%%%%%%%%%%%%%%%%%%%%%%%%%%%%
\section{Introduction}

%generalised entropies
Generalised entropy functions have been studied intensively
in the second half of the past century. They have been called
quasi-entropies in \cite {PD86}. Every entropy function
is in fact minus a relative entropy, also called a divergence. It is
relative to some reference measure $c$.
Consider the f-divergence \cite {CI72}
\be
I(p||c)=\sum_ac_af(p_a/c_a),
\label {intro:fdiv}
\ee
with $f(u)$ a convex function defined for $u>0$ and strictly convex at $u=1$.
It is minus the entropy of $p$, relative to $c$. 
Taking $c_a=1$ for all $a$ and $f(u)=u\ln u$ one obtains the Boltzmann-Gibbs-Shannon
entropy
\be
I(p)=-\sum_ap_a\ln p_a.
\label {intro:BGS}
\ee
Note that throughout the paper discrete probabilities are considered,
with events $a$ belonging to a finite or countable alphabet $A$.

Recent interest in these generalised entropies within statistical physics
goes back to the introduction by Tsallis \cite {TC88} of the $q$-entropy
\be
I_q(p)=\frac 1{1-q}\left(\sum_ap_a^q-1\right),
\label {intro:tsallisentropy}
\ee
with $q>0$. In the limit $q=1$ it converges to (\ref {intro:BGS}).
It has been studied before in the mathematics literature by Havrda and Charvat \cite {HC67},
and by Dar\'oczy \cite {DZ70}. Investigations within the physics community have lead
to some interesting developments. One of them is the introduction of deformed
logarithmic and exponential functions \cite {TC94,NJ02} --- see the Section 13.
They have been very useful to generalise common concepts,
like that of an exponential family or of a Gaussian distribution.
They also helped to clarify the pitfalls of the generalisation process.
One of the surprises is the necessity to introduce escort
probability functions \cite {TMP98} --- see Section 11.
In a series of papers, including \cite {NJ04c,NJ04d}, the present author
has elaborated a formalism based on deformed logarithms. In the present work,
it is shown that slightly more general results are obtained when abandoning
these deformed logarithms.

%survey of the paper
In Sections 2 to 6 the maximum entropy principle and the variational principle
are discussed in the context of generalised entropies. In particular, a characterisation
of the maximising probability distributions is given. This is used in Section 7 to
define a generalised exponential family. In Section 8 it is shown that
the intersection of distinct generalised exponential families is empty
and that there exists a one-to one-relation with generalised entropy
functions. Sections 9 tot 12 discuss geometric aspects, starting
with concepts from thermodynamics and introducing escort families
and a generalised Fisher information matrix.  Sections 13 and 14
discuss non-extensive thermostatistics and the percolation problem
as examples of the generalised formalism.
The paper ends with a short diascussion in Section 15.

%%%%%%%%%%%%%%%%%%%%%%%%%%%%%%%%%%%%%%%%%%%%%%%%%%%%%
\section{Generalised entropies}

%notations
Let us fix some further notations. The space of probability
distributions is denoted ${\cal M}_1^+(A)$.
Expectation values are denoted
$\langle p,X\rangle=\sum_{a\in A}p_aX(a)$.
Here we follow the physics tradition to put the elements of the dual space at the l.h.s..

It is rather common to define a generalised entropy
as any function $I(p)$ of the form
\be
I(p)=\sum_{a\in A}h\left(p_a\right),
\label {genent:entdef}
\ee
where $h(u)$ is a continuous strictly concave function,
defined on $[0,1]$, which vanishes when $u=0$ or $u=1$.
This is a special case of minus the f-divergence (\ref {intro:fdiv}), with weights $c_a=1$.
The entropy function $I(p)$ is defined for any $p\in {\cal M}_1^+(A)$ and has
values in $[0,+\infty]$.
In the present paper it is allowed that the function $h(u)$ is stochastic, this means,
depends also on $a$ in $A$.
But for convenience of notation, this dependence will not be made explicit.

Throughout the paper it is assumed that the derivative
\be
\frac {{\rm d}h}{{\rm d}u}=-f(u)
\label {genent:hder}
\ee
exists on the interval $(0,1)$ and defines a continuous function on the
halfopen interval $(0,1]$.
Because $h(u)$ is strictly concave, $f(u)$ is strictly increasing.
Note that it is allowed to diverge to $-\infty$ at $u=0$.
This is indeed the case when $h(u)=-u\ln u$ and $f(u)=1+\ln u$.

The function $f(u)$ can be used to rewrite the entropy $I(p)$ as
\be
I(p)=\sum_{a\in A}\int^1_{p_a}{\rm d}u\,f(u)=-\sum_{a\in A}\int_0^{p_a}{\rm d}u\,f(u)
=-\sum_{a\in A}p_a\int_0^1{\rm d}v\,f(p_av).
\label {genent:entrewr}
\ee
Note that the latter expression implies that
\be
I(p)\ge -\sum_{a\in A}p_af(p_a).
\ee

The standard definition of the Bregman divergence \cite {BLM67} reads
\be
D(p||q)=I(q)-I(p)-\sum_{a\in A}(p_a-q_a)f(q_a).
\label {genent:bregman}
\ee
In the case that $f(u)$ diverges at $u=0$ it is only well defined
when $q_a=0$ implies $p_a=0$.
It is a convex function of the first argument.
Note that one can write
\be
D(p||q)=\sum_{a\in A}\int_{q_a}^{p_a}{\rm d}u\,\left[f(u)-f(q_a)\right].
\ee
From the latter expression it is immediately clear that $D(p||q)\ge 0$,
with equality if and only if $p=q$.

%%%%%%%%%%%%%%%%%%%%%%%%%%%%%%%%%%%%%%%%%%%%%%%%%%%%%
\section{Maximum entropy principle}

Let be given a finite number of real functions $H_1(a)$, $H_2(a)$, $\cdots$, $H_n(a)$.
Assume they are bounded from below. In a physical context these functions may be
called Hamiltonians.
The maximum entropy problem deals with finding the probability
distribution $p$ that maximises $I(p)$ under the constraint that the expectation values
of the Hamiltonians $H_j$ attain given values $U_j$, called energies. Introduce the notation
\be
{\cal P}_U=\{p\in {\cal M}_1^+:\,\langle p,H_j\rangle=U_j,j=1,2,\cdots,n\}.
\label {maxent:constr}
\ee
Then one looks for the probability distribution $p\in {\cal P}_U$ which maximises $I(p)$.

\begin{definition}
A probability distribution $p^*\in {\cal P}_U$ is said to satisfy the
maximum entropy principle if it satisfies
\be
I(p)\le I(p^*)<+\infty\quad\mbox{ for all }p\in {\cal P}_U.
\ee
\end{definition}

In what follows a stronger condition is needed.
It was introduced some 40 years ago
\cite {RD67} --- see Theorem 7.4.1 of \cite {RD69} ---
and is in fact a stability criterion.
\begin{definition}
A probability distribution $p^*$ is said to satisfy the
variational principle if there exist parameters $\theta_1,\theta_2,\cdots,\theta_n$
such that
\be
+\infty>I(p^*)-\sum_{j=1}^n\theta_j\langle p^*,H_j\rangle
\ge I(p)-\sum_{j=1}^n\theta_j\langle p,H_j\rangle\quad\mbox{ for all }p\in {\cal M}_1^+.
\label {maxent:varprin}
\ee
\end{definition}
In statistical physics, a probability distribution satisfying the variational principle
is called an equilibrium state.

%%%%%%%%%%%%%%%%%%%%%%%%%%%%%%%%%%%%%%%%%%%%%%%%%%%%%
\section{Lagrange multipliers}

A popular way to solve the maximum entropy problem is by the introduction of
Lagrange parameters. However, a difficulty arises, known as the cutoff problem.
It is indeed possible that some of the probabilities $p_a$ of the optimising
probability distribution vanish. Let us see how this problem arises.
The Lagrangean reads
\be
{\cal L}=I(p)-\alpha\sum_{a\in A}p_a-\sum_{j=1}^n\theta_j\langle p,H_j\rangle.
\ee
Here, $\alpha$ is the parameter introduced to fix the normalisation condition
$\sum_{a\in A}$ $p_a=1$, the $\theta_j$ are introduced to cope with the constraints
(\ref {maxent:constr}). Variation of $\cal L$ w.r.t.~the $p_a$ yields
\be
f(p_a)=-\alpha-\sum_{j=1}^n\theta_j H_j(a).
\label {maxent:eqpdf}
\ee
The problem that can arise is that
it may well happen that the r.h.s.~of this expression does not belong
to the range of the function $f(u)$. This situation is particularly likely to occur when
$f(u)$ does not tend to $-\infty$ when $u$ tends to 0.
If the r.h.s.~is in the range of $f(u)$ then $p_a$ is determined uniquely by (\ref {maxent:eqpdf})
because of the assumption that $f(u)$ is a strictly increasing function.

The above problem is well known in optimisation theory. Because the constraints, defining
${\cal P}_U$, are affine, the set ${\cal P}_U$ forms a simplex. Its faces are obtained
by putting some of the probabilities $p_a$ equal to zero.
Because the entropy function $I(p)$ is concave it attains its maximum within one of these faces.
This observation leads to the ansatz that the probability distribution $p$,
which maximises $I(p)$ with $p$ in ${\cal P}_U$, if it exists, is determined by
a subset $A_0=\{a\in A:\,p_a=0\}$, and by the values of the parameters $\alpha$ and $\theta_j$,
which determine the remaining probabilities via (\ref {maxent:eqpdf}).
Let us now try to prove this statement.

%%%%%%%%%%%%%%%%%%%%%%%%%%%%%%%%%%%%%%%%%%%%%%%%%%%%%
\section{Characterisation}

Let us first consider the more familiar situation that $f(0)=-\infty$.

\begin{lemma}
Assume $f(0)=-\infty$.
Let $p^*\in {\cal M}_1^+$ satisfy the variational principle.
Then $p^*_a>0$ holds for all $a\in A$.
\end{lemma}

\beginproof

The inverted statement is proved.

Because of the normalisation, there exists at least one $a\in A$ for which $p^*_a>0$.
Assume $b\in A$ such that $p^*_b=0$. Let us show that
this implies that $p^*$ does not satisfy the variational principle.

Fix $0<\epsilon<<1$.
Introduce a new probability distribution $p$ which coincides with $p^*$ except that
\be
p_a=(1-\epsilon)p^*_a
\quad\mbox{ and }\quad p_b=\epsilon p^*_a.
\ee
Let
\be
M(\epsilon)=I(p)-\sum_{j=1}^n\theta_j\langle p,H_j\rangle.
\ee
Then one has
\be
\frac {{\rm d}M}{{\rm d}\epsilon}=f((1-\epsilon)p^*_a)-f(\epsilon p^*_a)
-\sum_{j=1}^n\theta_jp^*_a\left[H_j(a)-H_j(b)\right].
\ee
From the assumption $f(0)=-\infty$ then follows that
\be
\lim_{\epsilon\downarrow 0}\frac {{\rm d}M}{{\rm d}\epsilon}=+\infty.
\ee
This proves that $p^*$ does not satisfy the variational principle
because for $\epsilon$ sufficiently small $M(\epsilon)$ is strictly
larger than $M(0)$.

\endproof

\begin{theorem}
Assume $f(0)=-\infty$. A probability distribution $p^*$ satisfies the variational principle
if and only if there exists $\alpha$ and $\theta_1$, $\theta_2$, $\cdots$, $\theta_n$
such that (\ref {maxent:eqpdf}) holds for all $a\in A$.
\end{theorem}

\beginproof
First assume that $p^*$ satisfies (\ref {maxent:eqpdf}). This implies that $p^*_a>0$
for all $a\in A$ because $f(0)$ is not defined. Hence, the
divergence $D(p||p^*)$ is well defined for all $p$. Next one calculates
\be
D(p||p^*)&=&I(p^*)-I(p)-\sum_{a\in A}(p_a-p^*_a)f(p^*_a)\crcr
&=&I(p^*)-I(p)-\sum_{a\in A}(p_a-p^*_a)\left[-\alpha-\sum_{j=1}^n\theta_j H_j(a)\right]\crcr
&=&I(p^*)-I(p)+\sum_{j=1}^n\theta_j\langle p-p^*, H_j\rangle.
\label {char:div}
\ee
Because $D(p||p^*)\ge 0$ with equality if and only if $p=p^*$
there follows that $p^*$ satisfies the variational principle.

Next assume that $p^*$ satisfies the variational principle (\ref {maxent:varprin}).
From the lemma then follows that $p^*_a>0$ for all $a\in A$.
Hence, the divergence $D(p||p^*)$ is well-defined for all $p\in {\cal M}_1^+$.
It follows from the variational principle that
\be
D(p||p^*)&=&I(p^*)-I(p)-\sum_{a\in A}(p_a-p^*_a)f(p^*_a)\crcr
&\ge&\sum_{j=1}^n\theta_j\langle p^*-p,H_j\rangle-\sum_{a\in A}(p_a-p^*_a)f(p^*_a).
\label {theorem:proof1}
\ee
Now, the function $p\rightarrow D(p||p^*)$ is convex with continuous derivatives.
The r.h.s.~of the above expression is affine. Both l.h.s.~and r.h.s.~vanish
for $p=p^*$. One then concludes that the r.h.s.~is tangent to the convex function
and must be identically zero. One concludes that for all $p$
\be
\sum_{a\in A}(p_a-p^*_a)f(p^*_a)=\sum_{j=1}^n\theta_j\langle p^*-p,H_j\rangle.
\label {theorem:proof0}
\ee
This implies that $f(p^*_a)$ is of the form (\ref {maxent:eqpdf}) ---
take $p_a=\delta_{a,b}$ for some fixed $b$ to see this.

\endproof

%%%%%%%%%%%%%%%%%%%%%%%%%%%%%%%%%%%%%%%%%%%%%%%%%%%%%
\section{The case with cutoff}

Assume now that $f(0)=\lim_{u\downarrow 0}f(u)$ converges.
Then the divergence $D(p||q)$ is well defined for any pair of
probability distributions $p$, $q$.

\begin{theorem}
Assume that $f(0)=\lim_{u\downarrow 0}f(u)$ converges.
Are equivalent
\begin{enumerate}
\item {} $p^*$ satisfies the variational principle;
\item {} there exist parameters $\alpha$ and $\theta_1$, $\theta_2$, $\cdots$, $\theta_n$,
and a subset $A_0$ of $A$ such that
\begin {itemize}
\item {} (\ref {maxent:eqpdf}) is satisfied for all $a\in A\setminus A_0$;
\item {} $\displaystyle p^*_a=0\quad\mbox{ for all }a\in A_0$;
\item {} $\displaystyle
f(0)+\sum_{j=1}^n\theta_jH_j(a)\ge -\alpha
\quad\mbox{ for all }a\in A_0$.
\end{itemize}
\end{enumerate}
\end{theorem}

Note that this last condition expresses that the r.h.s.~of (\ref {maxent:eqpdf}) is out of the range of $f(u)$
because it takes a value less than $f(0)$.

\beginproof

\paragraph {1) implies 2)}
As in the proof of the previous Theorem, one shows that (\ref {theorem:proof1}) holds for all $p$.
But now one cannot conclude (\ref {theorem:proof0}) because some of the $p^*_a$ may vanish
so that $p^*$ lies in one of the faces of the simplex ${\cal M}_1^+$.
But one can still derive (\ref {maxent:eqpdf}) for all $a$ for which $p^*_a\not=0$.

Assume now that $p^*_a=0$ for some given $a\in A$. Let
\be
p_b=(1-\epsilon)p^*_b+\epsilon\delta_{b,a}.
\ee
Then the l.h.s.~of (\ref {theorem:proof1}) becomes
\be
D(p||p^*)&=&\sum_{b\in A}^{\not=a}\int^{p^*_b}_{(1-\epsilon)p^*_b}{\rm d}u\,\left[f(p^*_b)-f(u)\right]
+\int_0^{\epsilon}{\rm d}u\,\left[f(u)-f(0)\right]\crcr
&\le&\epsilon\sum_{b\in A}p^*_b\left[f(p^*_b)-f((1-\epsilon)p^*_b)\right]
+\int_0^{\epsilon}{\rm d}u\,f(u)-\epsilon f(0)\crcr
&=&\mbox{O}(\epsilon^2).
\ee
On the other hand, the r.h.s.~of (\ref {theorem:proof1}) becomes
\be
\mbox{r.h.s.}&=&
\epsilon\sum_{j=1}^n\theta_j\sum_{b\in A}p^*(b)H_j(b)
-\epsilon\sum_{j=1}^n\theta_jH_j(a)
+\epsilon\sum_{b\in A}p^*_bf(p^*_b)
-\epsilon f(0).
\ee
From the inequality (\ref {theorem:proof1}) then follows
\be
0\ge \sum_{j=1}^n\theta_j\langle p^*,H_j\rangle-\sum_{j=1}^n\theta_jH_j(a)
+\sum_{b\in A}p^*_bf(p^*_b)
-f(0).
\ee
This implies the desired inequality because
\be
-\alpha=\sum_{b\in A}p^*_bf(p^*_b)+\sum_{j=1}^n\theta_j\langle p^*,H_j\rangle.
\ee

\paragraph {2) implies 1)}

One calculates
\be
I(p)-\sum_{j=1}^n\theta_j\langle p,H_j\rangle
&=&-D(p||p^*)+I(p^*)-\sum_{a\in A}(p_a-p^*_a)f(p_a^*)
-\sum_{j=1}^n\theta_j\langle p,H_j\rangle\crcr
&\le&I(p^*)-f(0)\sum_{a\in A_0}p_a\crcr
& &
+\sum_{a\in A\setminus A_0}(p_a-p^*_a)\left[\alpha+\sum_{j=1}^n\theta_jH_j(a)\right]
-\sum_{j=1}^n\theta_j\langle p,H_j\rangle\crcr
&=&I(p^*)-\sum_{j=1}^n\theta_j\langle p^*,H_j\rangle\crcr
& &
-\sum_{a\in A_0}p_a\left[f(0)+\alpha
+\sum_{j=1}^n\theta_jH_j(a)\right].
\label {cutoff:temp1}
\ee
The variational principle now follows using the third assumption of the Theorem.

\endproof

%%%%%%%%%%%%%%%%%%%%%%%%%%%%%%%%%%%%%%%%%%%%%%%%%%%%%
\section{Statistical models}

In the definition of the variational principle
there is given a set of Hamiltonians $H_1(a)$, $H_2(a)$, $\cdots$, $H_n(a)$,
this means, real functions over the alphabet $A$, bounded from below.
The equilibrium distribution $p^*$ is then characterised by a normalisation
constant $\alpha$, by parameters $\theta_1$, $\theta_2$, $\cdots$, $\theta_n$,
and by a subset $A_0$ of the alphabet $A$ --- see (\ref {maxent:eqpdf}).
The emphasis now shifts towards these parameters.

\begin{theorem}
Let be given Hamiltonians $H_1(a)$, $H_2(a)$, $\cdots$, $H_n(a)$.
For each $\theta$ in $\Ro^n$ there exists at most one probability distribution
$p^*$ satisfying the variational principle (\ref {maxent:varprin}) with these
parameters $\theta$.
\end{theorem}

\beginproof

If $p^*$ and $q^*$ both satisfy the variational principle (\ref {maxent:varprin}) with
the same parameters $\theta$ then also the convex combination
$r^*=\frac 12p^*+\frac 12q^*$ has the same property because the entropy function
is concave. But then one can conclude from the inequalities (\ref {maxent:varprin})
that $I(r^*)=\frac 12I(p^*)+\frac 12I(q^*)$. Because the entropy function
is strictly concave there follows $p^*=q^*$.

\endproof

The set of $\theta$ for which a $p^*$ exists,
satisfying the variational principle (\ref {maxent:varprin}),
is denoted $\cal D$. The probability distribution is denoted $p_\theta$
instead of $p^*$. The constant $\alpha$ appearing in (\ref {maxent:eqpdf})
is replaced by $\alpha(\theta)$.

A statistical model is a parametrised set of probability distributions.
The above Theorem implies that the set $(p_\theta)_{\theta\in{\cal D}}$,
 of probability distributions satisfying the variational principle,
is a statistical model. One can say that such a model belongs to the
generalised exponential family.

\begin {definition}
\label {statmod:defexpfam}
Let be given a generalised entropy function $I(p)$ of the form (\ref {genent:entdef}).
A statistical model $(p_\theta)_{\theta\in{\cal D}}$ belongs to the
generalised exponential family if there exist real functions $H_1(a)$, $H_2(a)$, $\cdots$, $H_n(a)$,
bounded from below, such that each member $p_\theta$ of the model satisfies
the variational principle (\ref {maxent:varprin}) with these Hamiltonians and with this
set of parameters.
\end {definition}

Clearly, entropy functions which differ only by a scalar factor determine the same
generalised exponential family.

%%%%%%%%%%%%%%%%%%%%%%%%%%%%%%%%%%%%%%%%%%%%%%%%%%%%%
\section{Uniqueness theorem}

Let us now turn to the question whether a given model $(p_\theta)_{\theta\in{\cal D}}$
can belong to two different generalised exponential families.

\begin {theorem}
\label {uniq:thm}
Let be given a model $(p_\theta)_{\theta\in{\cal D}}$.
Assume that there exists an open subset ${\cal D}_0$ of $\cal D$ with the property
that the set of values of $p_{\theta,a}$ covers the open interval $(0,1)$
\be
(0,1)\subset\{p_{\theta,a}:\,\theta\in {\cal D}_0, a\in A\}.
\ee
If the model belongs to two different generalised exponential families, one with
entropy function $I_1(p)$, the other with entropy function $I_2(p)$,
then there exists a constant $\lambda$ such that $I_2(p)=\lambda I_1(p)$
for all $p$.
\end {theorem}

\beginproof
Take any point $u$ in $(0,1)$ and a corresponding $\theta\in {\cal D}_0$ and $a$ such that $p_{\theta,a}=u$.
From the previous theorems follows that there exist functions $\alpha_i(\theta)$
and Hamiltonians  $H_{i1}(a)$, $H_{i2}(a)$, $\cdots$, $H_{in}(a)$,
with $i=1,2$, such that
\be
p_{\theta,a}=f^{-1}_{i,a}\left(-\alpha_i(\theta)-\sum_{j=1}^n\theta_jH_{i,j}(a)\right).
\ee
Let $F_a=f_{2,a}\circ f^{-1}_{1,a}$. Note that this is a strictly increasing continuous function.
Then one has
\be
F_a\left(-\alpha_1(\theta)-\sum_{j=1}^n\theta_jH_{1,j}(a)\right)=
-\alpha_2(\theta)-\sum_{j=1}^n\theta_jH_{2,j}(a).
\label {uniq:temp}
\ee
This relation holds also on a vicinity of $\theta\in {\cal D}_0$.
It therefore implies the existence of $\lambda_a$ and $K_{i,j}$ such that
\be
H_{2,j}(a)-K_{2,j}=\lambda_a (H_{1,j}(a)-K_{1,j}),
\quad j=1,2,\cdots,n.
\ee
Then one can rewrite (\ref {uniq:temp}) as
\be
F_a(v)=\gamma_a(\theta)+\lambda_a v,
\ee
with
\be
\gamma_a(\theta)=-\alpha_2(\theta)-\sum_{j=1}^n\theta_j K_{2,j}
+\lambda_a\left[\alpha_1(\theta)+\sum_{j=1}^n\theta_j K_{1,j}\right],
\label {uniq:temp2}
\ee
valid for some neighbourhood of the given $\theta$.
Using the definition of $F_a(v)$ one obtains
\be
f_{2,a}(u)&=&\gamma_a(\theta)+\lambda_af_{1,a}(u),
\ee
valid on some neighbourhood of the given $u\in (0,1)$.
Because $u$ is arbitrary and the functions $f_{ia}$ are continuous, the same expression must hold on all of $(0,1]$.
From $0=h_{i,a}(0)=\int_0^1{\rm d}u\,f_{i,a}(u)$ now follows that $\gamma_a(\theta)=0$.
Therefore (\ref {uniq:temp2}) becomes
\be
\lambda_a=\frac {\alpha_2(\theta)+\sum_{j=1}^n\theta_j K_{2,j}}{\alpha_1(\theta)+\sum_{j=1}^n\theta_j K_{1,j}}.
\ee
In particular, $\lambda_a$ does not depend on $a\in A$.
One concludes therefore that there exists $\lambda$ so that $f_{2,a}(u)=\lambda f_{1,a}(u)$.
This implies $I_2(p)=\lambda I_1(p)$.

\endproof

%%%%%%%%%%%%%%%%%%%%%%%%%%%%%%%%%%%%%%%%%%%%%%%%%%%%%
\section{Thermodynamics}

Throughout this Section, let be given a statistical model
$(p_\theta)_{\theta\in{\cal D}}$ belonging to the
generalised exponential family.

Note that if $p_\theta$ and $p_\eta$ both belong to the same set ${\cal P}_U$
%(see Section \ref {sect:maxent})
then they satisfy $I(p_\theta)=I(p_\eta)$.
Hence, a function $S(U)$ can be defined by
\be
S(U)=I(p_\theta)\quad\mbox{ whenever }\langle p_\theta,H_j\rangle=U_j
\mbox{ for }j=1,2,\cdots,n.
\label {therm:sudef}
\ee
This function is called the thermodynamic entropy.
The concept of thermodynamic entropy was first introduced by Clausius around 1850.
The Legendre transform of the thermodynamic entropy is given by
\be
\Phi(\theta)=\sup\{S(U)-\sum_{j=1}^n\theta_jU_j\}.
\ee
This function was introduced by Massieu in 1869.
The suprememum is taken over all $U$ for which $S(U)$ is defined by (\ref {therm:sudef}).
The function is convex --- this is a well-known property of Legendre transforms.

\begin {proposition}
\label {therm:proprel}
One has
\be
\Phi(\theta)=I(p_\theta)-\sum_{j=1}^n\theta_j\langle p_\theta,H_j\rangle,
\qquad\theta\in {\cal D}.
\ee
\end {proposition}

\beginproof
Given $\theta\in {\cal D}$ there exists $p_\theta$ for which the variational principle holds.
Then one has, with $U_j=\langle p_\theta,H_j\rangle$,
\be
I(p_\theta)-\sum_{j=1}^n\theta_j\langle p_\theta,H_j\rangle
=S(U)-\sum_{j=1}^n\theta_jU_j\le \Phi(\theta).
\ee
This proves the inequality in one direction.
Next, fix $\epsilon>0$ and let $U$ be such that
\be
\Phi(\theta)\le S(U)-\sum_{j=1}^n\theta_jU_j+\epsilon,
\ee
with $U$ such that $S(U)$ is defined by (\ref {therm:sudef}).
Then, there follows from the definition of $S(U)$ that $\eta\in{\cal D}$ exists
such that $S(U)=I(p_\eta)$ with $\langle p_\eta,H_j\rangle=U_j$, $j=1,2,\cdots,n$.
The variational principle now implies that
\be
I(p_\theta)-\sum_{j=1}^n\theta_j\langle p_\theta,H_j\rangle
&\ge&
I(p_\eta)-\sum_{j=1}^n\theta_j\langle p_\eta,H_j\rangle\crcr
&=&S(U)-\sum_{j=1}^n\theta_jU_j\crcr
&\ge& \Phi(\theta)-\epsilon.
\ee
Because $\epsilon>0$ is arbitrary, the inequality in the other direction follows now.

\endproof

The inverse Legendre transformation reads
\be
\overline S(U)=\inf_\theta\{\Phi(\theta)+\sum_{j=1}^n\theta_jU_j\}.
\ee
It is a concave function.

\begin {proposition}
One has $S(U)=\overline S(U)$ for all $U$ for which $S(U)$ is defined by (\ref {therm:sudef}).
\end {proposition}

\beginproof

From the definition of the Massieu function $\Phi(\theta)$ there follows that
\be
\Phi(\theta)\ge S(U)-\sum_{j=1}^n\theta_jU_j
\quad\mbox{ for all }\theta\in\Ro^n.
\ee
This implies that $S(U)\le \overline S(U)$.
On the other hand, from the definition (\ref {therm:sudef}) of $S(U)$ follows that
\be
S(U)=\Phi(\theta)+\sum_{j=1}^n\theta_jU_j,
\ee
where $\theta$ is such that $p_\theta\in{\cal P}_U$. This implies $S(U)\ge \overline S(U)$.
The two inequalities together establish the desired equality.

\endproof

%%%%%%%%%%%%%%%%%%%%%%%%%%%%%%%%%%%%%%%%%%%%%%%%%%%%%
\section{Thermodynamic relations}

Like in the previous Section, there is given a statistical model
$(p_\theta)_{\theta\in{\cal D}}$ belonging to the
generalised exponential family. In addition, let ${\cal D}_0$ be an open subset
of $\cal D$ on which the map $\theta\rightarrow \langle p_\theta,H_j\rangle$
is continuous.

The following results are typical properties of Legendre transforms.
For completeness, proofs are given.

\begin {proposition}
\label {thermrel:phiderprop}
The first derivative of the Massieu function $\Phi(\theta)$ exists for $\theta$ in ${\cal D}_0$.
It satisfies
\be
\frac {\partial\Phi}{\partial\theta_j}=-\langle p_\theta,H_j\rangle,
\quad \theta\in {\cal D}_0.
\label {thermrel:fd}
\ee
\end {proposition}

\beginproof
From the definitions one has for $\theta$ and $\theta+\eta$ in ${\cal D}_0$
\be
\Phi(\theta+\eta)&=&I( p_{\theta+\eta})-\sum_{j=1}^n(\theta_j+\eta_j)\langle  p_{\theta+\eta},H_j\rangle\crcr
&\ge&I( p_{\theta})-\sum_{j=1}^n(\theta_j+\eta_j)\langle  p_{\theta},H_j\rangle\crcr
&=& \Phi(\theta)-\sum_{j=1}^n\eta_j\langle p_\theta,H_j\rangle,
\ee
and
\be
\Phi(\theta)&=&I(p_\theta)-\sum_{j=1}^n\theta_j\langle p_\theta,H_j\rangle\crcr
&\ge&I(p_{\theta+\eta})-\sum_{j=1}^n\theta_j\langle p_{\theta+\eta},H_j\rangle\crcr
&=&\Phi(\theta+\eta)+\sum_{j=1}^n\eta_j\langle p_{\theta+\eta},H_j\rangle.
\ee
Expression (\ref {thermrel:fd}) now follows
using the continuity of the map $\theta\rightarrow \langle p_\theta,H_j\rangle$.

\endproof

Introduce the metric tensor
\be
g_{i,j}(\theta)=
\frac {\partial^2\Phi}{\partial\theta_i\partial\theta_j}.
\label {thermrel:fdd}
\ee
Because the Massieu function $\Phi(\theta)$ is convex
the matrix $g(\theta)$ is positive definite, whenever it exists.
By the previous Proposition one has
\be
g_{i,j}(\theta)=-\frac {\partial\,}{\partial\theta_i}
\langle p_\theta,H_j\rangle
\label {thermrel:energder}
\ee
for those $\theta$ in ${\cal D}_0$ for which the derivative exists.

In thermodynamics, the derivative of $S(U)$ equals the inverse of the absolute temperature $T$.
Here, the analogous property becomes

\begin {proposition}
Let $\theta\in{\cal D}_0$ and define $U$ by $U_j=\langle p_\theta,H_j\rangle$.
Then one has
\be
 \frac {\partial S}{\partial U_j}=\theta_j,
\quad j=1,2,\cdots,n.
\label {thermrel:entder}
\ee

\end {proposition}

\beginproof

On a vicinity of $\theta$ is $S(U)=\Phi(\theta)+\sum_{j=1}^n\theta_jU_j$.
Hence, one can write
\be
\frac {\partial S}{\partial\theta_j}
&=&\sum_{k=1}^n\left(\frac {\partial\Phi}{\partial\theta_k}+U_k\right)\frac {\partial\theta_k}{\partial U_j}+\theta_j.
\ee
But the first term in the r.h.s.~vanishes because the previous Proposition holds.
Hence, the desired result follows.

\endproof

The two relations (\ref {thermrel:fd}) and (\ref {thermrel:entder})
are dual in the sense of Amari \cite {AS85}. In thermodynamics, the entropy $S(U)$ and
Massieu's function $\Phi(\theta)$ are state functions, the energies $U_j$ are extensive thermodynamic
variables, the parameters $\theta_j$ are the intensive thermodynamic variables.

%%%%%%%%%%%%%%%%%%%%%%%%%%%%%%%%%%%%%%%%%%%%%%%%%%%%%
\section{Escort probabilities}

Let us now make the additional assumption that the function $f(u)$,
which enters the definition (\ref {genent:entrewr}) of the generalised entropy,
has a derivative $f'(u)$. Because $f(u)$ was supposed to be strictly increasing,
one can write
\be
f(u)=f(1)-\int^1_u{\rm d}v\,\frac 1{\phi(v)},
\quad u\in (0,1],
\label {escort:phidef}
\ee
where $\phi(v)=1/({\rm d}f/{\rm d}v)$ is a strictly positive function.

As before, there is given a statistical model
$(p_\theta)_{\theta\in{\cal D}}$ belonging to the
generalised exponential family, and ${\cal D}_0$ is an open subset
of $\cal D$ on which the map $\theta\rightarrow \langle p_\theta,H_j\rangle$
is continuous. The set $A_0(\theta)$ is the set of $a\in A$ for which $p_\theta(a)=0$.
From theorems 1 and 2 now follows
\be
\frac {\partial\,}{\partial\theta_j}p_{\theta,a}=\phi(p_{\theta,a})
\left(-\frac {\partial\alpha}{\partial\theta_j}-H_j(a)\right),
\quad \theta\in {\cal D}_0,a\in A\setminus A_0(\theta).
\label {escort:pder}
\ee
This expression was used in \cite {NJ04d} as a condition under which
a generalisation of the well-known bound of Cram\'er and Rao is optimal.
An immediate consequence of (\ref {escort:pder}) is

\begin {proposition}
\label {escort:prop3}
Assume the regularity condition
\be
0=\sum_a\frac {\partial\,}{\partial\theta_j}p_{\theta}(a).
\label {escort:regular}
\ee
Assume in addition that
\be
z(\theta)=\sum\strut'\phi(p_{\theta,a})<+\infty,
\label {escort:partsum}
\ee
where $\sum\strut'$ denotes the sum over all $a\in A\setminus A_0(\theta)$.
Then one has
\be
\frac {\partial\alpha}{\partial\theta_j}=-\frac 1{z(\theta)}\sum\strut'\phi(p_{\theta,a})H_j(a).
\label {escort:probder}
\ee
\end {proposition}

\beginproof
On a vicinity of the given $\theta$ one has (\ref {escort:pder}).
Hence, by summing (\ref {escort:pder}) over $a\in A\setminus A_0(\theta)$ one obtains
using (\ref {escort:regular})
\be
0&=&\sum\strut'\phi(p_{\theta,a})
\left(-\frac {\partial\alpha}{\partial\theta_j}-H_j(a)\right),
\quad \theta\in {\cal D}_0,a\in A\setminus A_0(\theta).
\ee
\endproof

The probability distribution
\be
P_{\theta,a}&=&\frac 1{z(\theta)}\phi(p_{\theta,a}),
\quad p_{\theta,a}\not=0,\crcr
&=&0, \qquad\mbox{ otherwise},
\label {escort:escort}
\ee
when it exists, is called the escort of the exponential family $(p_\theta)_{\theta\in{\cal D}}$.
With this notation, one can write the result of the Proposition as
\be
\frac {\partial\alpha}{\partial\theta_j}=-\langle P_{\theta},H_j\rangle.
\label {escort:id}
\ee

%%%%%%%%%%%%%%%%%%%%%%%%%%%%%%%%%%%%%%%%%%%%%%%%%%%%%
\section{Generalised Fisher information}

Let be given a model $(p_\theta)_{\theta\in{\cal D}}$ for which
$z(\theta)$, as given by (\ref {escort:partsum}), converges.
The escort probabilities $P_{\theta,a}$ are defined by (\ref {escort:escort}).
Then one can define a generalised Fisher information matrix by
\be
I_{i,j}(\theta)&=&\langle P_\theta,X_i(\theta)X_j(\theta)\rangle,
\label {fish:fishdef}
\ee
where the score variables are defined by
\be
X_{i,a}(\theta)\equiv\frac 1{P_{\theta,a}}\frac{\partial\,}{\partial\theta_i}p_{\theta,a}.
\ee
Note that in the standard case of $h(u)=-u\ln u$ one has $\phi(u)=u$ so that
the escort probabilities $P_\theta$ coincide with the $p_\theta$.
Then (\ref {fish:fishdef}) reduces to the conventional definition.

Fix now a set of Hamiltonians $H_1(a)$, $H_2(a)$, $\cdots$, $H_n(a)$.
Then one can define a covariance matrix $\sigma(\theta)$ by
\be
\sigma_{i,j}(\theta)=\langle P_\theta,H_iH_j\rangle-\langle P_\theta,H_i\rangle\,\langle P_\theta,H_j\rangle.
\ee

\begin {proposition}
Assume a finite alphabet $A$. Then one has
\be
I_{i,j}(\theta)=z(\theta)g_{i,j}=z^2(\theta)\sigma_{i,j}.
\label {genfish:result}
\ee

\end {proposition}

\beginproof
From (\ref {escort:pder}) follows
\be
X_{j,a}(\theta)&=&z(\theta)\left(-\frac {\partial\alpha}{\partial\theta_j}-H_j(a)\right)
\ee
for all $\theta\in {\cal D}_0$ and $a\in A\setminus A_0(\theta)$.
Hence, the Fisher information matrix becomes
\be
I_{i,j}(\theta)&=&z^2(\theta)\sum_{a\in A}P_{\theta,a}
\left(-\frac {\partial\alpha}{\partial\theta_i}-H_i(a)\right)
\left(-\frac {\partial\alpha}{\partial\theta_j}-H_j(a)\right).
\ee
Using (\ref {escort:id}) there follows $I_{i,j}(\theta)=z^2(\theta)\sigma_{i,j}$.

On the other hand, from (\ref {thermrel:energder}) and (\ref {escort:pder}) there follows
\be
g_{i,j}(\theta)
&=&-\frac {\partial\,}{\partial\theta_i}\sum_{a\in A}p_{\theta,a}H_j(a)\crcr
&=&-\sum_{a\in A}P_{\theta,a}
\left(-\frac {\partial\alpha}{\partial\theta_i}-H_i(a)\right)H_j(a).
\label {fisher:prop4calc}
\ee
Using (\ref {escort:probder}) there follows $g_{i,j}(\theta)=z(\theta)\sigma_{i,j}$.
\endproof

The assumption of a finite alphabet is made to ensure that
the conditions of Proposition \ref {escort:prop3} are fulfilled and that the sum
and derivative may be interchanged in (\ref {fisher:prop4calc}).

The generalised inequality of Cram\'er and Rao, in the present notations,
reads \cite {NJ04d}
\be
\left(\sum_{kl}\sigma_{kl}u_ku_l\right)
\left(\sum_{kl}I_{kl}v_lv_k\right)
\ge \left(\sum_{kl}g_{kl}u_kv_l\right)^2,
\ee
with $u$ and $v$ arbitrary real vectors.
The previous Proposition then implies that the inequality
becomes an equality when $u=v$, when $P$ is related to $p$ via (\ref {escort:escort}),
and when $p_\theta$ belongs to a generalised exponential family.

%%%%%%%%%%%%%%%%%%%%%%%%%%%%%%%%%%%%%%%%%%%%%%%%%%%%%
\section{Non-extensive thermostatistics}

Define the $q$-deformed logarithm by \cite {TC94,TC04}
\be
\ln_q(u)=\frac 1{1-q}\left(u^{1-q}-1\right).
\ee
It is a strictly increasing function, defined for $u>0$.
Indeed, its derivative equals
\be
\frac {{\rm d}\,}{{\rm d}u}\ln_q(u)=\frac 1{u^q}>0.
\ee
In the limit $q=1$ the $q$-deformed logarithm converges to the nature logarithm $\ln u$.

The deformed logarithm can be used in more than one way to define an entropy function.
The $q$-entropy (\ref {intro:tsallisentropy}) can be written as
\be
I_q(p)=\sum_{a\in A}p_a\ln_q\left(\frac 1{p_a}\right).
\label {tsallis:entropy}
\ee
Comparison with (\ref {genent:entdef}) gives
\be
h(u)=\frac u{1-q}\left(u^{q-1}-1\right)=u\ln_q\left(\frac 1{u}\right).
\ee
One has $h(0)=h(1)=0$. Taking the derivative gives
\be
f(u)=-\frac {{\rm d}h}{{\rm d}u}=\frac 1{q-1}\left(qu^{q-1}-1\right).
\ee
It is a strictly increasing function on $(0,1]$ when $q>0$.
The function $\phi(u)$ is given by
\be
\phi(u)=\frac 1qu^{2-q}.
\label {ne:phi}
\ee
The probability distributions belonging to the generalised exponential
family, corresponding with (\ref {tsallis:entropy}), are
\be
p_a=q^{1/(1-q)}\left[1-(q-1)\alpha-(q-1)\sum_j\theta_jH_j(a)\right]_+^{1/(q-1)},
\ee
with $[u]_+=\max\{0,u\}$.
This is indeed the kind of probability distribution discussed in the original
paper of Tsallis \cite {TC88}. However, more often used is the alternative
of \cite {TMP98}. In the latter paper
the concept of escort probability distributions was introduced into the literature.
They were defined by
\be
P_a=\frac 1Zp_a^q,
\ee
which in the present notations corresponds with $\phi(u)$ proportional to $u^q$.
This can be obtained by replacing the constant $q$ by $2-q$ in (\ref {tsallis:entropy}).
The entropy function then reads
\be
I(p)=-\sum_ap_a\ln_q(p_a),
\ee
which is \emph {not} the expression that one would write down based
on the information theoretical argument that $\ln(1/p_a)$ is
the amount of information (counted in units of $\ln 2$),
gained from an event occurring with probability $p_a$.
Note that with this definition of entropy function the condition $q<2$
is needed in order to satisfy the requirements that 
the function $f(u)=\frac {{\rm d}\,}{{\rm d}u}(u\ln_q (u)$ is an increasing function.

%%%%%%%%%%%%%%%%%%%%%%%%%%%%%%%%%%%%%%%%%%%%%%%%%%%%%
\section{The percolation problem}

This example has been treated in \cite {NJ05_78}. It is a
genuine example of an important model of statistical physics
which does not belong to the exponential family. In addition,
it is an example which fits into the present generalised
context provided that one allows that the function $h(u)$
appearing in the definition (\ref {genent:entdef}) of the generalised entropy
function is stochastic.

In the site percolation problem \cite {SD85}, the points of a lattice are occupied
with probability $q$, independent of each other. The point at the
origin is either unoccupied, with probability $p_\emptyset$, or it belongs
to a cluster of shape $i$, with probability $p_i$. This cluster is finite
with probability 1, provided that $0\le q\le q_c$, where $q_c$ is the percolation
threshold. The probability $p_\infty$ that the origin belongs to
an infinite cluster is strictly positive for $q>q_c$.
However, for the sake of simplicity of the presentation, $0<q<q_c$ will be assumed
--- see \cite {NJ05_78} for the general case.

These probabilities are given by
\be
p_i=c_iq^{s(i)}(1-q)^{t(i)},
\label {cases:percdist}
\ee
where $c_i$ is the number of different clusters of shape $i$, $s(i)$
is the number of occupied sites in the cluster, and $t(i)$ is the number
of perimeter sites, this is, of unoccupied neighbouring sites.
Note that (\ref {cases:percdist}) also holds when the
origin is not occupied, provided that one convenes that
$c(\emptyset)=1$, $s(\emptyset)=0$ and $t(\emptyset)=1$.

Choose the Hamiltonian
\be
H(i)=\frac {t(i)}{t(i)+s(i)}.
\label {cases:percham}
\ee
and introduce the parameter $\theta$ by
\be
\theta=\ln\frac {q}{1-q},
\qquad
q=\frac 1{1+e^{-\theta}}.
\ee
Then one can write
\be
\ln \frac {p_i}{c_i}=\left[-\alpha(\theta)-\theta H(i)\right]\,\left[s(i)+t(i)\right],
\label {cases:prer}
\ee
with
\be
\alpha(\theta)=\ln(1+e^{-\theta})
\ee
This looks like an exponential family, except for the extra factor $[s(i)+t(i)]$
in the r.h.s..
Introduce the stochastic function
\be
f_i(u)&=&\frac {\ln u}{s(i)+t(i)}.
\ee
Then the above expression is of the form (\ref {maxent:eqpdf}).
By integrating $f_i(u)$ one obtains
\be
h_i(u)&=&-\frac {u\ln u}{s(i)+t(i)}.
\ee
It is now straightforward to verify that the percolation problem belongs to
a generalised exponential family.
The relevant entropy function for the percolation model in the non-percolating region $0<q<q_c$ is therefore
\be
I(p)=-\sum_i\frac {p_i\ln p_i}{s(i)+t(i)}.
\ee

%%%%%%%%%%%%%%%%%%%%%%%%%%%%%%%%%%%%%%%%%%%%%%%%%%%%%
\section{Discussion}

%summary
Sections 3 to 6 of the present paper discuss the variational principle,
which is stronger than the maximum entropy principle. It is shown that
the method of Lagrange multipliers leads to the correct result, even in the
context of generalised entropy functions. The difficulty that arises
is known as the cutoff problem: the optimising probability distribution
may assign vanishing probabilities to some of the events. To cope
with this situation the two cases have been considered separately.
Theorem 1 treats the standard case,
Theorem 2 copes with the vanishing probabilities.

In Section 7, a generalised definition of an exponential
family is given. It identifies the members of the generalised exponential
family with the solutions of the variational principle, given
a generalised entropy function of the usual form (\ref {genent:entdef}).
The definition of the standard exponential family corresponds of course with the
Boltzmann-Gibbs-Shannon entropy. Entropy functions $I(p)$ and $\lambda I(p)$,
with $\lambda>0$, determine the same exponential family.
Assuming some technical condition, 
the intersection of different generalised exponential families is empty
--- see Theorem \ref {uniq:thm}.
As a consequence, a one-to one relation has been established between
generalised exponential families and classes of equivalent entropy
functions.

In \cite {NJ04d}, the notion of phi-exponential family was introduced.
The 'phi' in this name
refers to the function $\phi(v)$, introduced in (\ref {escort:phidef}).
It is one divided by the derivative of the function $f(v)$ appearing in the
expression (\ref {genent:entrewr}) for the entropy function $I(p)$.
The assumption that the derivative of $f(v)$ exists for all $v>0$
has been eliminated in the present paper. More important is that the
definition of a generalised exponential family is now given directly
in terms of the entropy function $I(p)$, via the variational principle,
without relying on the notion of deformed exponential functions.

Sections 9 to 12 discuss the geometric properties of a generalised
exponential family, using a terminology coming from 150 year old
thermodynamics. The main result is (\ref {genfish:result}),
proving the equality of the three quantities
generalised Fisher information, metric tensor times partition sum $z(\theta)$,
and covariance matrix multiplied with $z^2(\theta)$.
The covariance matrix is calculated using the
escort family of probability distributions.

%applications
Many applications of generalised exponential families are found in the literature,
in the context of nonextensive thermostatistics.
The latter has been discussed in Section 13. A completely different
kind of example is found in percolation theory --- see Section 14.
It illustrates the possibility that the function $f(u)$, which determines the
entropy function $I(p)$ via (\ref {genent:entrewr}), is of a stochastic nature.
One can expect that many other applications will be found in the near future.

%further generalisations
Finally note that the generalisation of the present work to quantum probabilities
is straightforward. Let be given a strictly increasing function $f(u)$,
continuous on $(0,1]$. The expression (\ref {genent:entrewr}) can be generalised
to
\be
I(\rho)=-\int_0^1{\rm d}v\,\Tr \rho f(v\rho),
\ee
where $\rho$ is any density operator in a Hilbert space.
The Bregman divergence (\ref {genent:bregman}) generalises to
\be
D(\rho||\rho')=I(\rho')-I(\rho)-\Tr(\rho-\rho')f(\rho).
\ee
The basic inequality $D(\rho||\rho')\ge 0$ is proved using Klein's inequality
--- see 2.5.2.~of \cite {RD69}.

%%%%%%%%%%%%%%%%%%%%%%%%%%%%%%%%%%%%%%%%%%%%%%%%%%%%%%%%%%%%
\section*{Acknowledgements}
This work has benefitted from a series of discussions with Flemming Tops\o e.

\bibliographystyle{plain}
\makeatletter
\renewcommand\@biblabel[1]{#1. }
\makeatother
\bibliography{naudts}

\begin{thebibliography}{10}

\bibitem{AS85}
S.~Amari.
\newblock {\em Differential-geometrical methods in statistics}, volume~28 of
  {\em Lecture Notes in Statistics}.
\newblock Springer, New York, Berlin, 1985.

\bibitem{BLM67}
L.M. Bregman.
\newblock The relaxation method of finding a common point of convex sets and
  its application to the solution of problems in convex programming.
\newblock {\em USSR Comp. Math. Math. Phys.}, 7:200--217, 1967.

\bibitem{CI72}
I.~Csisz\'ar.
\newblock A class of measures of informativity of observation channels.
\newblock {\em Per. Math. Hung.}, 2:191--213, 1972.

\bibitem{DZ70}
Z.~Dar\'oczy.
\newblock {\em Inform. Control}, 16:36, 1970.

\bibitem{HC67}
J.~Havrda and F.~Charvat.
\newblock {\em Kybernetica}, 3:30, 1967.

\bibitem{NJ02}
J~Naudts.
\newblock Deformed exponentials and logarithms in generalized thermostatistics.
\newblock {\em Physica A}, 316:323--334, 2002.

\bibitem{NJ04c}
J.~Naudts.
\newblock Continuity of a class of entropies and relative entropies.
\newblock {\em Rev. Math. Phys.}, 16:809--822, 2004.

\bibitem{NJ04d}
J.~Naudts.
\newblock Estimators, escort probabilities, and phi-exponential families in
  statistical physics.
\newblock {\em J. Ineq. Pure Appl. Math.}, 5:102, 2004.

\bibitem{NJ05_78}
J.~Naudts.
\newblock Parameter estimation in nonextensive thermostatistics.
\newblock {\em Physica A}, 365:42--49, 2006.

\bibitem{PD86}
D.~Petz.
\newblock Quasi-entropies for finite quantum systems.
\newblock {\em Rep. Math. Phys.}, 23:57--65, 1986.

\bibitem{RD67}
D.~Ruelle.
\newblock A variational formulation of equilibrium statistical mechanics and
  the gibbs phase rule.
\newblock {\em Commun. Math. Phys.}, 5:324--329, 1967.

\bibitem{RD69}
D.~Ruelle.
\newblock {\em Statistical mechanics}.
\newblock W.A. Benjamin, New York, 1969.

\bibitem{SD85}
D.~Stauffer.
\newblock {\em Introduction to percolation theory}.
\newblock Plenum Press, New York, 1985.

\bibitem{TC88}
C.~Tsallis.
\newblock Possible generalization of boltzmann-gibbs statistics.
\newblock {\em J. Stat. Phys.}, 52:479--487, 1988.

\bibitem{TC94}
C.~Tsallis.
\newblock What are the numbers that experiments provide?
\newblock {\em Quimica Nova}, 17:468, 1994.

\bibitem{TC04}
C.~Tsallis.
\newblock Nonextensive statistical mechanics: construction and physical
  interpretation.
\newblock In M.~Gell-Mann and C.~Tsallis, editors, {\em Nonextensive Entropy},
  pages 1--53, Oxford, 2004. Oxford University Press.

\bibitem{TMP98}
C.~Tsallis, R.S. Mendes, and A.R. Plastino.
\newblock The role of constraints within generalized nonextensive statistics.
\newblock {\em Physica A}, 261:543--554, 1998.

\end{thebibliography}

%%%%% Correction made on April 17, 2007 %%%%%
%\vspace{12pt}\noindent \copyright \ 2007 by MDPI (http://www.mdpi.org). Reproduction is permitted for noncommercial purposes.
%%% Original
%%% \vspace{12pt}\noindent \copyright \ 2006 by MDPI (http://www.mdpi.org). Reproduction is permitted for noncommercial purposes.

\end{document}